\documentclass[aoas]{imsart}

\RequirePackage{amsthm,amsmath,amsfonts,amssymb}
\RequirePackage[authoryear]{natbib}
\RequirePackage[colorlinks,citecolor=blue,urlcolor=blue]{hyperref}
\RequirePackage{graphicx}

\RequirePackage{lineno}
\usepackage{tabularx}
\usepackage{float}
\usepackage[table]{xcolor}
\usepackage{ulem}

\startlocaldefs

\def\vec#1{\mathchoice{\mbox{$\mathbf\displaystyle\bf#1$}}
{\mbox{$\mathbf\textstyle\bf#1$}}
{\mbox{$\mathbf\scriptstyle\bf#1$}}
{\mbox{$\mathbf\scriptscriptstyle\bf#1$}}}

\endlocaldefs

\begin{document}

\begin{frontmatter}

\title{Modeling Homophily in Exponential-Family Random Graph Models for Bipartite Networks}
\runtitle{Homophily for Bipartite ERGMs}

\begin{aug}
\author[A]{\fnms{Rashmi P}~\snm{Bomiriya}\ead[label=e1]{rashmi.bomiriya@gmail.com}},
\author[B]{\fnms{Alina R}~\snm{Kuvelkar}\ead[label=e2]{ark336@psu.edu}}
\author[B]{\fnms{David R}~\snm{Hunter}\ead[label=e3]{drh20@psu.edu}}
\and
\author[C]{\fnms{Steffen}~\snm{Triebel}\ead[label=e4]{S.Triebel@exeter.ac.uk}}
\address[A]{R S Metrics Asia Holdings Pvt Limited\printead[presep={,\ }]{e1}}

\address[B]{Pennsylvania State University\printead[presep={,\ }]{e2,e3}}

\address[C]{University of Exeter Business School\printead[presep={,\ }]{e4}}
\end{aug}

\begin{abstract}
Homophily, the tendency of individuals who are alike to form ties with one another,
is an important concept in the study of social networks.
Yet accounting for homophily effects is complicated in the context of bipartite networks where
ties connect individuals not with one another but rather with a separate
set of nodes,
which might also be individuals but which are often an entirely different type of objects.
As a result, much work on the effect of homophily in a bipartite network proceeds by first
eliminating the bipartite structure, collapsing a two-mode network to a one-mode network
and thereby ignoring potentially meaningful structure in the data.
We introduce a set of methods to model homophily on bipartite networks without
losing information in this way, then we demonstrate
that these methods allow for substantively interesting findings in management science
not possible using standard techniques.
These methods are implemented in the widely-used \texttt{ergm} package for \texttt{R}.
\end{abstract}


\end{frontmatter}

\section{Introduction and Motivation}
\label{intro}
Networks are an increasingly popular method for capturing patterns of interactions among 
pairs, or more generally sets, of individuals.  In its most basic form, a network consists
of a fixed set of nodes, also called vertices,
along with a set of pairs of those nodes.  Each pair is called a tie,
or an edge, and may be thought of as connecting the two nodes by a particular
relationship whose definition depends on the application.
Relationships in networks are often thought to be influenced by partner selection mechanisms. 
One of the most popular partner selection mechanisms is called homophily, the basic idea of which
is that ties are more likely to form between individuals who match on one or 
more attributes. This partner selection mechanism then regularly leads to the formation of 
homogeneous groups. The importance of homophily is well-established both in academia, as is 
evident by the literature in different fields of the social sciences
\citep{mcpherson2001,ertug2022},
as well as outside of academia, as is reflected by the well-known proverb ``birds of a feather flock together.''
It is no surprise, then, that its incorporation into statistical models for social networks is often straightforward.

Yet dealing with homophily is not straightforward for bipartite networks, also known as 
affiliation or two-mode networks.  Technically,
a bipartite network consists of two disjoint sets, or modes, of vertices, $V_1$ and $V_2$,
and a set of edges $E\subseteq V_1 \times V_2$.  That is,
every edge must link a vertex from mode 1 to a vertex from mode 2.
in layman's terms, ``actors'' comprising one mode of a bipartite network are affiliated with 
``events'' of its second mode, so connections among actors arise only indirectly
through joint affiliations to events and
vice versa. As there are multiple types of two-mode networks where the terms ``actor'' and ``event''
do not accurately describe vertices, we simply refer to them as ``mode 1'' and ``mode 2''.
In cases in which bipartite networks are composed of modes of the same node type---for instance,
in a heterosexual contact network where partnerships only occur between opposite-gender
nodes---homophily involving, say, age category or race may be considered as in the standard,
non-bipartite case.
Such a composition, however, is rare. Typically, the two modes of nodes are
comprised of entirely different types of entities, and homophily is only theoretically meaningful 
among entities of the same type. It is these latter cases in which the current article is relevant.

As an example of the potential complexity of homophily in two-mode networks,
consider the case of corporate boards, in which directors (mode 2)
are affiliated with companies (mode 1). 
Interlocking directorates, which occur ``when a person affiliated with one organization sits on the 
board of directors of another organization'' \citep{mizruchi1996}, are studied by scholars 
from various disciplines, such as sociology \citep{chu2016} or management research 
\citep{lamb2016,martin2015}. 
Here, researchers may take on a resource 
dependency perspective \citep{pfeffer2003} and consequently raise questions about the role of 
industry homophily, estimating whether
directors are more likely to sit on the boards of firms that belong to 
the same industry rather than distinct industries. Yet different conceptions
of how to measure the strength of homophily can give rise to different statistics.
It is a fundamental theoretical difference if 
homophily is estimated 
(a) by counting through how many directors a focal firm is connected to another homophilous firm,
or (b) by counting to how many different homophilous firms a director connects a focal firm.
These perspectives represent a node-centric and an edge-centric view, respectively.
If these counts are totaled for the entire network, the two views produce equivalent 
homophily statistics, and a straightforward approach
does not distinguish between them.

This paper introduces a method to put either the node- or edge-centric perspective into 
focus and estimate homophily via a statistic that is nonlinear in, respectively, the number of
paths between a given homophilous node pair or the number of edges that complete paths with
homophilous endpoints starting with a given edge.  This method 
allows for a deeper integration of social science theories and network models while
providing mathematical flexibility that can lead to improved modeling performance.

We present two applications that illustrate the relevance of this method
to various disciplines prone to 
social network analysis, such as sociology, management research, and social psychology. Our 
results suggest that the reformulated homophily effects we introduce to the analysis of two-mode 
networks allow researchers from various fields to find nuanced answers to contemporary questions.

The first application 
is a dataset collected among members of an organization in which subjects listed their core 
competencies.
We show that using the node-centric reformulation
not only improves model fit as measured by log-likelihood,
but produces estimates that allow for meaningful interpretation
regarding the distribution of hard and soft skills (such as financial analysis versus planning) in an organization, adding to ongoing scholarly discussions and creating implications for practitioners \citep{laker2011,robles2012}.

The second application refers to the aforementioned directorates. Analyzing the 
network of boards of directors among the Fortune 500, we argue that in this dataset, the role of 
gender is best captured by making use of the edge-centric reformulation of two-mode 
homophily. 
Our results 
indicate that in addition to the proverbial glass ceiling, i.e., the difficulty faced by females in
ascending the corporate ladder as evidenced by the fact that only 32\% of board members are
female, there may be a ``glass door''
that moderates females' horizontal integration into the corporate elite. These findings add to 
ongoing debates about the role of demographic attributes in social elite cohesion and corporate 
governance \citep{zhu2014, mcdonald2013, park2013}.

Both of these applications are examined in Section~\ref{sec:application}.
First, we establish the statistical foundation, which is based on a modeling framework
known as exponential-family random graph models, or ERGMs.  Section~\ref{sec:ergm}
discusses particular complications of modeling relationships in a bipartite network and
describes the ERGM framework.  Section~\ref{sec:weightedNodematch} 
introduces novel modeling tools within the ERGM framework, explaining their favorable
theoretical properties from both a statistical and application-specific perspective.
Section~\ref{sec:discussion} 
offers some concluding remarks, while Appendices~\ref{sec:changestats}
and~\ref{sec:curvedEF} provide additional theoretical material regarding the
use of these methods in an ERGM framework.

\section{Bipartite Networks and Exponential-Family Random Graph Models}
\label{sec:ergm}

Basic notions of analyzing bipartite networks are reviewed by 
\citet{borgatti1997}, 
who explain how many familiar network techniques and measurements---e.g.,
visualization of networks, nodal degree, network density, betweenness and centrality---may be extended 
to the bipartite case. 
\citet{latapy2008basic} cite numerous applications of two-mode networks and
similarly extend additional common notions,
introducing statistics like the ``redundancy coefficient'' or expanding on others such as the 
bipartite clustering coefficient of 
\citet{robins2004}.  

When the modeling goal is to understand relationships among 
nodes belonging to just one of the two modes, as is the case when studying homophily,
a commonly-used technique is to project
from a two-mode network onto either of the two modes.  Here, we explain projections and
introduce the modeling framework of exponential-family random graph models, or ERGMs.

\subsection{Projections onto one-mode networks}
\begin{figure}[tb]
  \begin{center}
  \includegraphics[width=4in, height=1.5in]{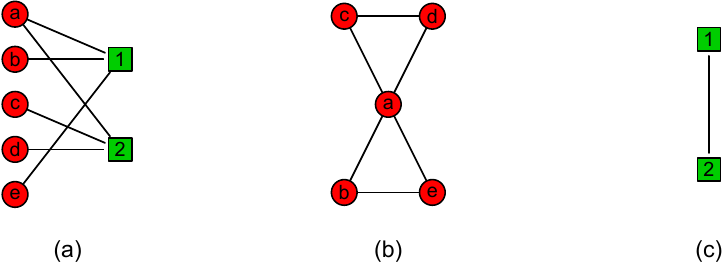}
  \caption{Networks (b) and (c) depict the projections of the original 
  two-mode network (a) onto mode 1 
  nodes (circles) and mode 2 nodes (squares), respectively. An edge between 
  a pair of nodes exists in the projection if and only if there is at least one two-path 
  between those two nodes in (a).}
  \label{fig:projection}
  \end{center}
\end{figure}

Any two-mode network gives rise to two separate one-mode projections, as illustrated by
Figure~\ref{fig:projection}.  The edges in the projections can include weights, indicating the
number of distinct nodes of the opposite mode that connect to a given node pair. 
To consider homophily, we might apply standard modeling techniques,
such as those presented in Section~\ref{subsec:ergm}, to either of the projections.

Yet as 
\citet{latapy2008basic} 
point out, the projection approach is not ideal for several reasons:  In brief, 
the projection loses information about the original bipartite structure of the network,
and it may even 
distort certain types of information such as clustering coefficients.
For instance, in the interlocking directorates example cited in Section~\ref{intro},
projecting the underlying two-mode 
structure onto one-mode networks that depict connections between 
firms \citep{withers2020, howard2017} loses information about attributes on the director level is 
almost lost in its entirety.  Similarly,
while there are various studies about the role of director-level attributes, 
the network context in which corporate directors maneuver is typically neglected 
\citep{westphal2007, mcdonald2013}.

For bipartite networks, we let $n=n_1+n_2$, 
where nodes 1 through $n_1$ are of mode~1 and nodes $n_1+1$ through $n$ are of mode~2.
We denote a random network by 
the $n \times n$ matrix $\vec Y$,
where $Y_{ij}$ equals 0 or 1 according to whether the $(i,j)$th edge is absent or present.
In this case, $\vec Y$ may only have nonzero entries in the off-diagonal rectangular
submatrix in rows 1 to $n_1$ and columns $n_1+1$ to $n$, along with its mirror image since
$\vec Y$ is usually symmetric in a bipartite setting.  The matrix $\vec Y^2$ contains both projection networks
from Figure~\ref{fig:projection}(b) and (c), with the mode~1 projection in the upper-left $n_1\times n_1$
submatrix and the mode~2 projection in the lower-right $n_2\times n_2$ submatrix.  
Each entry in these submatrices gives the number of distinct paths of length two connecting
the two nodes indexed by the row and column numbers of the entry; in particular, nonzero
entries of $\vec Y^2$ indicate edges in one of the projection networks.

\subsection{Exponential-family Random Graph Models}
\label{subsec:ergm}

The aim of this article is to model the nonzero elements of $\vec Y$ itself rather than $\vec Y^2$,
and for this purpose we turn to the family of probability models known as exponential-family random graph models (ERGMs).
\citet{robins2007ait} give a general introduction to these
models, which are sometimes called p-star models in reference to the seminal paper of
\citet{wasserman1996lma}.
These models define a parametric family of probability distributions on the space of all possible networks.
The basic ERGM may be written
\begin{equation}\label{basicERGM}
P_\theta (\vec Y = \vec y) = \frac{\exp\{ \sum_{i=1}^p \theta_i s_i(\vec y) \}}{\kappa(\vec\theta)},
\quad \vec y \in {\cal Y},
\end{equation}
where $s_1(\vec y), \ldots, s_p(\vec y)$ are user-defined statistics measured on the
network $\vec y$ and we denote the vector of all network statistics by $\vec s(\vec y)$. 
When covariates $X$ should be included in the model, we may add $X$ to the notation
and write $\vec s(\vec y, X)$, where we allow these statistics to depend on any available known
covariates \citep{hunter2008ergm}. 
The parameters $\theta_1, \ldots, \theta_p$ are the 
corresponding unknown coefficients to be 
estimated, ${\cal Y}$ is the set of all allowable networks,
and $\kappa(\vec\theta)$ is a normalizer necessary to ensure that 
Equation~(\ref{basicERGM}) defines a legitimate probability distribution.
Several authors have explored the use of ERGMs in the context of bipartite networks.  For instance,
\citet{wang2016} review the use of ERGMs in the context of a more general network paradigm that includes
two-mode networks as a special case.  \citet{kevork2022} explore the use of ERGMs to model bipartite networks
in the case where nodal covariates are themselves considered random, rather than fixed and known as in the
current article.

In this framework, 
homophily related to variable X may be woven into a model for one-mode networks
by ensuring that the
sufficient statistics that define the model include the count of all ties between two 
individuals who match on X.  
An illustrative homophily-based model for mutual friendships among a group of people that assumes
all undirected edges form independently with some probability $p_1$ between people of the opposite gender
and with some different probability $p_2$ between people of the same gender can be written
as
\begin{equation}\label{homophilyERGM}
P(\vec Y = \vec y) \propto \exp\left \{ \theta_1 \sum_{i < j} y_{ij}  +
\theta_2 \sum_{i<j} y_{ij} I\{ \mbox{$i$ and $j$ are of the same gender} \}  \right \} .
\end{equation}
Equation~(\ref{homophilyERGM}) defines $s_1(\vec y)$ and $s_2(\vec y)$ as the number of edges in
$\vec y$ and the number of edges with nodes that match on gender, respectively.
One may show 
that under model~(\ref{homophilyERGM}), $p_1$ and $p_2$ are given by
$e^{\theta_1}/(1+e^{\theta_1})$ and $e^{\theta_1+\theta_2}/(1+e^{\theta_1+\theta_2})$, respectively.

Using the \texttt{ergm} package for \texttt{R} \citep{r2023}, the homophily statistic $s_2(\vec y)$
may be incorporated into an ERGM for one-mode networks by adding
\texttt{nodematch("gender")} to the model specification.  The
\texttt{ergm} package, which implements the methods described in this article,
is part of the \texttt{statnet} suite of packages described in detail in 
volume 24 of the {\it Journal of Statistical Software} 
\citep[e.g.,][]{handcock2008, hunter2008ergm}. More recent enhancements of the \texttt{ergm}
package are given by \citet{krivitsky2023}.

By contrast, an analogous method of measuring homophily in the context of bipartite graphs
would include a statistic counting two-paths---i.e., paths of 
length two---connecting nodes of the same category,
since it is impossible that such nodes are connected directly.
Authors such as 
\citet{faust2002} and \citet{FHH2004} 
incorporate bipartite homophily effects
in ERGMs in this way, by considering two-paths classified according to the
categories of their endpoints. 
Yet it may not be desirable to count every two-path with 
matching endpoints, for two reasons.  

First is a law of diminishing returns:  If two nodes are already
connected via a two-path, every additional two-path connecting them might be
less important than the previous one, so a linear relationship between the number of
two-paths and the homophily statistic may be inappropriate from a modeling perspective.
A method to model this type of diminishing effect is via the alternating $k$-twopath 
or alternating $k$-star statistics of \citet{robins2007rdi}.  These statistics are reformulated
as geometrically weighted degree and geometrically weighted dyadic shared partner statistics,
respectively, by \citet{hunter2007}.  As we explain in Appendix~\ref{sec:curvedEF}, the latter of these
two statistics and the new methods we propose are closely related.

The second reason relates to the fact that homophily in a bipartite network, because
it is intrinsically a function of multiple edges, destroys the independence among edges implied
by models like Equation~(\ref{homophilyERGM}) that implement homophily in one-mode networks.
Lack of dependence can sometimes result in a model that exhibits degeneracy, an issue described by
\citet{handcock2003} and \citet{schweinberger2011}.  
In brief, degeneracy can result because 
it is possible, by adding a single edge to a network in a particular configuration, to increase the
number of two-stars (say) by a very large number, up to the number of first- or second-mode nodes.
Since other statistics cannot compensate for this large increase in such situations, degenerate models
often put inappropriately high probability mass on networks with very large or very small numbers of 
edges, depending on whether the coefficient of the two-star term is positive or negative, even with 
coefficient values that are in some sense optimal.

\section{Modeling homophily using ERGMs}
\label{sec:weightedNodematch}

The approach we outline here involves a pair of homophily-based statistics that may be 
easily incorporated into an ERGM.  Our new statistics introduce two sliding scales,
at one end of which we find the fully linear default two-star statistic that can sometimes prove problematic.
The other end of the scale counts only the first two-star formed by each pair of matching nodes, 
or the first two-star (connecting matching nodes) that contains each edge, depending on 
the formulation of homophily we choose.  

Suppose we want to model the homophily effect of a particular categorical nodal attribute $c$
measured on the mode~1 nodes.  For example, if nodes represent people then $c_i$ could be the gender of node $i$.
We assume here only that the variable $c$ may be measured on 
mode~1 nodes; it need not even apply to mode~2 nodes.  A similar argument would apply in the
case where we wished instead to model a homophily effect for mode~2 nodes.  

As a starting point for measuring the degree of homophily, let us consider the 
analogue of the homophily statistic in Equation~(\ref{homophilyERGM}), namely,
the total number of 
two-paths that link one mode-1 node with another mode-1 node of the same 
category.  This statistic may be obtained by summing
$y_{ik}y_{jk}$ for all matching mode-1 nodes $i$ and $j$ and all mode-2 nodes $k$.  Thus,
we write the basic homophily statistic as
\begin{equation}\label{simple}
  \mbox{b1nodematch}(\vec y) = 
  \frac12 \sum_{i=1}^{n_1} \sum_{j=1}^{n_1} \sum_{k=n_1+1}^n 
  y_{ik}y_{jk} I\{ \mbox{$c_i=c_j$, $i\ne j$} \},
\end{equation}
where the ``b1'' in b1nodematch stands for ``bipartite of mode~1'' and $I\{ \cdot\}$ is the
indicator function taking either a zero or one depending on the falsity or truth
of the enclosed condition.  
We divide by 2 because the summation double-counts every relevant two-path.

\subsection{Edge-centered and node-centered views of homophily}

\begin{figure}[tb]
  \begin{center}
  \includegraphics[width=3.5in, height=1.5in]{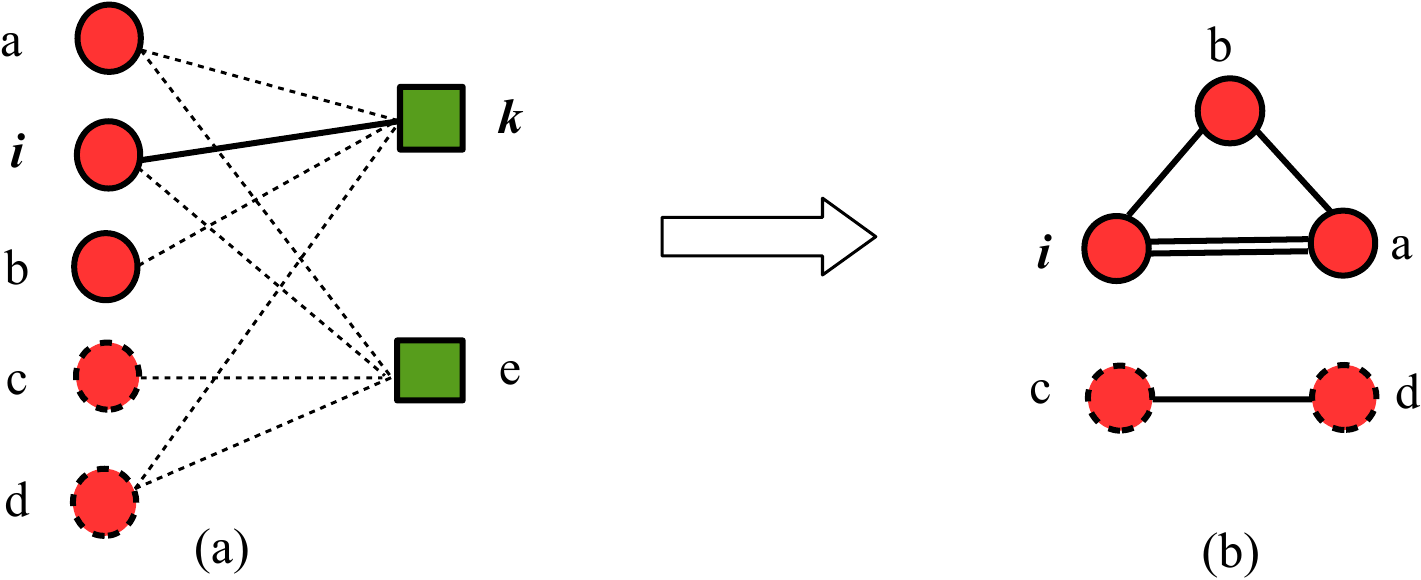}
  \caption{Mode-1 nodes are circles and mode-2 nodes are squares, 
  while dashed and continuous
  outlines indicate different levels of a mode-1 nodal attribute.  In (a), edge 
  $i\longleftrightarrow k$ 
  is part of two distinct two-paths joining matching mode-1 nodes,
  $a$---$k$---$i$ and $b$---$k$---$i$.
  In (b), we see the projection
  of (a) onto mode-1 nodes, with multiplicity indicated as appropriate.}
  \label{fig1}
  \end{center}
\end{figure}

A simple reformulation of Equation~(\ref{simple}) gives
\begin{equation}\label{rewriteEdgeCentered}
  \mbox{b1nodematch}(\vec y) = 
  \frac12
  \sum_{i=1}^{n_1} 
  \sum_{k=n_1+1}^{n}
  y_{ik}
  \left[ 
  \sum_{j \ne i} y_{jk}
  I\{ c_i=c_j \} 
  \right].
\end{equation}
Thus, for every edge $i\longleftrightarrow k$, i.e., whenever $y_{ik}=1$, we count the number of matching two-stars containing
this edge, sum, and finally divide by two since each two-star will have
been counted twice.
We illustrate this edge-centered view of the homophily statistic in
Figure~\ref{fig1}(a).  
There are two mode~1 nodes, labeled a and b, that match node $i$ (the categories are
denoted by the line style of the nodes' borders) that are also connected to $i$ through $k$.
Thus, the value in square brackets in Equation~(\ref{rewriteEdgeCentered}) is 2.  

By contrast, a different reformulation of Equation~(\ref{simple}) gives
\begin{equation}\label{rewriteNodepairCentered}
  \mbox{b1nodematch}(\vec y) = 
  \frac12
  \sum_{i=1}^{n_1}
   \sum_{\stackrel{ j\ne i :}{\mbox{\scriptsize $c_j=c_i$}}}
  \left[ 
 \sum_{k=n_1+1}^{n_1+n_2} y_{ik}y_{kj}
  \right],
\end{equation}
where the sum is taken over all pairs $i\ne j$ of matching mode~1 nodes; in this double sum, $(i,j)$ and $(j,i)$ are counted as distinct node pairs, so again
we divide by two to correct for the double-counting. 
We refer to Equation~(\ref{rewriteNodepairCentered}) as the node-centered view, 
because it counts, in the square brackets,
the number of two-paths that connect each pair $\{i,j\}$ of matching nodes.  
Using the network in Figure~\ref{fig1}(b), this count of two-paths equals 2 if $j=a$ and
1 if $j=b$.

Our modification to the two-star count involves modifying either 
Equation~(\ref{rewriteEdgeCentered}) or
Equation~(\ref{rewriteNodepairCentered}) by including an exponent 
in the range $[0,1]$.  In the node-centered view~(\ref{rewriteNodepairCentered}), we use 
the exponent $\alpha$, which results in
\begin{equation}
  \mbox{b1nodematch}(\vec y; \alpha) = \frac12
  \sum_{i=1}^{n_1}
   \sum_{\stackrel{ j\ne i :}{\mbox{\scriptsize $c_j=c_i$}}}
  \left[ 
   \sum_{k=n_1+1}^{n_1+n_2} y_{ik}y_{kj}
  \right]^\alpha
  \quad \mbox{with $0\leq \alpha \leq1$}. 
  \label{eqn:alpha}
\end{equation} 
On the other hand, if we take the edge-centered view~(\ref{rewriteEdgeCentered}),
we use the exponent $\beta$, which gives
\begin{equation}
  \mbox{b1nodematch}(\vec y; \beta) = \frac12
  \sum_{i=1}^{n_1}
  \sum_{k=n_1+1}^{n_1+n_2} y_{ik}
  \left[
  \mathop{\sum}_{\stackrel{ j\ne i :}{\mbox{\scriptsize $c_j=c_i$}}} y_{jk} 
  \right]^\beta 
  \quad\mbox{with $0 \leq \beta \leq1$}.
  \label{eqn:beta}
\end{equation}
When $\alpha=0$ or $\beta=0$
in Equations~(\ref{eqn:alpha}) and (\ref{eqn:beta}),
we interpret the expression
$0^0$ as zero, since the base will always be an integer whereas the exponent
can take values in $[0,1]$; thus,
$0^0$ is most sensibly interpreted as
$\lim_{\epsilon\to0} 0^\epsilon=0$.

Equations~(\ref{eqn:alpha}) and (\ref{eqn:beta}) may be modified slightly
if we wish to consider only nodes matching a certain value $v$.  This modification
yields the terms
\begin{equation}
  \mbox{b1nodematch}_v(\vec y; \alpha) = \frac12
  \sum_{i=1}^{n_1}
  \mathop{\sum}_{\stackrel{ j\ne i :}{\mbox{\scriptsize $c_j=c_i=v$}}}
  \left[ 
   \sum_{k=n_1+1}^{n_1+n_2} y_{ik}y_{kj}
  \right]^\alpha
  \label{eqn:alphaDifferential}
\end{equation} 
and
\begin{equation}
  \mbox{b1nodematch}_v(\vec y; \beta) = \frac12
  \sum_{i=1}^{n_1}
  \sum_{k=n_1+1}^{n_1+n_2} y_{ik}
  \left[
  \mathop{\sum}_{\stackrel{ j\ne i :}{\mbox{\scriptsize $c_j=c_i=v$}}} y_{jk} 
  \right]^\beta. 
  \label{eqn:betaDifferential}
\end{equation}
The versions of the statistics in Equations~(\ref{eqn:alpha}) and (\ref{eqn:beta}) model what is called ``uniform'' homophily, since
the homophily effect is implicitly presumed to be the same for all levels
of the categorical variable.  By contrast, Equations~(\ref{eqn:alphaDifferential}) and (\ref{eqn:betaDifferential}) model ``differential'' homophily
in which each level $v$ of the categorical variable corresponds to a separate statistic, and
not all levels' statistics need be included in the ERGM.

\subsection{Theoretical considerations}

When $\alpha=\beta=1$,
the two statistics of Equations~(\ref{eqn:alpha}) and (\ref{eqn:beta})
are equal to one another and to
the full count of two-stars with matching endpoints.
On the other end of the spectrum, when $\alpha=0$,
statistic~(\ref{eqn:alpha}) counts the number of matching node pairs connected by 
at least one two-path.  When $\beta=0$, statistic~(\ref{eqn:beta})  counts the number 
of edges involved in at least one matching two-star.
The case $\alpha=0$ is the only
case other than $\alpha=\beta=1$ where we know of an analogue 
elsewhere in the literature:
If the alternating $k$-twopath statistic of
\citet[][Equation 6.13]{wang2009} is extended to a matching-attribute-based alternating 
$k$-twopath statistic, then statistic~(\ref{eqn:alpha}) with $\alpha=0$ is equivalent 
to its special case of $\lambda_t=1$. However, we are not aware of any other 
direct correspondences between existing ERGM statistics and either 
of Equations~(\ref{eqn:alpha}) or ~(\ref{eqn:beta}).

The network statistics we have introduced are curved exponential family model terms in the sense of 
\citet{hunter2006iic}, which is a mathematically useful feature explained in Appendix~\ref{sec:curvedEF}.
In practical terms, this means that it should be possible to model $\vec\theta$ and
the discount parameter jointly,
rather than taking the profile likelihood approach that we have done here where the likelihood
is maximized only with respect to $\vec\theta$ for fixed $\alpha$ or $\beta$ on a grid of values in $[0,1]$.  
However, as of the writing of this manuscript, the {\tt ergm} package implementation of the {\tt b1nodematch} and {\tt b2nodematch}
model terms does not exploit {\tt ergm}'s full curved parameter capabilities.

Finally, models with the {\tt b1nodematch} or {\tt b2nodematch} terms can exhibit a mathematically unusual property.
Consider Model~(\ref{basicERGM}) with $p=1$ and $s_1(\vec y) = \mbox{b1nodematch}(\vec y; \alpha)$; that is,
\begin{equation}\label{b1nodematchmodel}
P (\vec Y = \vec y) = \frac{\exp\{  \theta \times \mbox{b1nodematch}(\vec y; \alpha) \}}{\kappa(\vec\theta)}
\end{equation}
for specific values $\theta$ and $\alpha$.  Then this ERGM is a dyadic dependence model in the sense that
the individual dyads---that is, the Bernoulli random variables $Y_{ik}$---are
not mutually independent.  Let $V=V_1\cup V_2$ denote the set of vertices, or nodes, decomposed into the
mode 1 vertices $V_1$ and the mode 2 vertices $V_2$.  For nonempty subsets $W_1\subset V_1$
and $W_2\subset V_2$, consider the possible bipartite networks on the nodes $W=W_1\cup W_2$, where
only edges connecting a node in $W_1$ with a node in $W_2$ are permitted.  According to a result of
\citet{shalizi2013}, Model~(\ref{b1nodematchmodel}) is not projective, which means that if we marginalize
over all networks on $V$ to get the marginal distribution for networks on $W$, this distribution is not the
same as Equation~(\ref{b1nodematchmodel}) for the set of networks on $W$ for all possible choices of $W_1$ and $W_2$.
However, Model~(\ref{b1nodematchmodel}) does enjoy a limited projectivity property: Projectivity
holds for the set of networks on any $W$ of the form $V_1\cup W_2$, where $W_2\subset V_2$.  This
property means, for instance, that standard properties such as consistency and
asymptotic normality of maximum likelihood estimators hold if we fix $n_1$ and
let $n_2$ tend to $\infty$.   These results hold if $\beta$ replaces $\alpha$ in Equation~(\ref{b1nodematchmodel})
or if additional dyadic independence statistics are added to the $\vec s(\vec y)$ vector.

\subsection{Interpreting Model Coefficients}
\label{sec:interpretation}

As a means for understanding the homophily statistics introduced in the previous section,
we first describe the notion of change statistics that is central to the theory of ERGMs.
Given a network $\vec y \in {\cal Y}$, 
the vector $\vec s(\vec y) = s_1(\vec y), \ldots, s_p(\vec y)$ 
of statistics from the ERGM of Equation~(\ref{basicERGM}) 
yields a set of change statistics that may be viewed as the off-diagonal entries of an $n\times n$ matrix
whose $(i,j)$ entry is defined as
\[
\vec \delta_{\vec s}(\vec y)_{ij} = \vec s(\vec y_{ij}^+) - \vec s(\vec y_{ij}^-),
\]
where $\vec y_{ij}^+$ and $\vec y_{ij}^-$ represent the networks obtained by 
fixing $y_{ij}=1$ or $y_{ij}=0$, respectively, and keeping all other entries
the same as in $\vec y$ itself.  These change statistics are important because
under the ERGM of Equation~(\ref{basicERGM}),
the conditional log-odds that $Y_{ij}=1$, conditional on all other entries in $\vec Y$
(denoted by $\vec Y_{ij}^c$), may be expressed as
\begin{equation}
\label{logOdds}
\log \frac{ P_{\vec\theta}(Y_{ij}=1 \mid \vec Y_{ij}^c=\vec y_{ij}^c) } 
{ P_{\vec\theta}(Y_{ij}=0 \mid \vec Y_{ij}^c=\vec y_{ij}^c)}
= \vec\theta^\top \vec \delta_{\vec s}(\vec y)_{ij}.
\end{equation}
Equation~(\ref{logOdds}) arises directly from Equation~(\ref{basicERGM}) via
simple algebra and omits the troublesome factor $\kappa(\vec\theta)$, 
a fact that is useful in myriad ways such as simulating random networks whose 
distribution is approximately $P_{\vec\theta}(\cdot)$ via MCMC
\citep[][Section 6]{hunter2008ergm}. 

Equation~(\ref{logOdds}) also reveals that, for $1\le \ell\le p$, 
the $\ell$th element $\theta_\ell (\vec \delta_{\vec s}(\vec y)_{ij})_\ell$ is the change in the log-odds of the entire network
resulting from changing the $i,j$ edge from 0 to 1, holding everything else equal.  For instance, in Equation~(\ref{homophilyERGM}), 
we see that $(\vec \delta_{\vec s}(\vec y)_{ij})_1=1$ and
$(\vec \delta_{\vec s}(\vec y)_{ij})_2=I\{\mbox{$i$ and $j$ are of the same gender}\}$ for all $i<j$, and these values do not depend on any of the values
in $\vec y_{ij}^c$.  Therefore, adding (subtracting) the edge $(i,j)$ increases
(decreases) the log-odds of the entire network by either $\theta_1$ or
$\theta_1+\theta_2$, depending on whether $i$ and $j$ are of the same gender,
independent of the rest of the network.
In Section~\ref{sec:application}, we
similarly interpret fitted models involving the homophily terms introduced
here.

\section{Substantive Findings in Two-Mode Network Datasets}
\label{sec:application}

In this section, we demonstrate that the proposed bipartite homophily statistics give rise to
meaningful results that allow for a more substantial integration with theoretical considerations
proposed by social scientists than the basic statistic of Equation~(\ref{simple}).
We do so by applying them to the two 
datasets introduced in the introduction: A network in which employees of an organization are 
affiliated with their core competencies and a network of directors that are affiliated with different 
firms. We contrast two approaches to using the reformulated homophily effects: First, a profile 
likelihood approach, which allows researchers to take a more explorative perspective that is 
relatively agnostic to prior theory, and second, a theory-based approach, that is relatively agnostic 
to changes in the log-likelihood.

\subsection{Competence network}
\label{sec:competence}

For the competence network discussed in Section~\ref{intro},
$n_1=176$ actors were asked to list up to three core competencies each 
from a list of $n_2=33$ competencies. 
Competencies included a variety of topics, such as financial analysis, information technology, 
marketing in the context 
of different countries, and industrial sales. 
Social scientists researching such a network may be interested in the role of hard skill 
competencies which refer to concrete abilities such as having knowledge about a manufacturing 
process and are nested in the individual, versus competencies that are more integrated into the 
organizational structure, such as the capability for strategic planning or privileges for commodities 
procurement. 
Understanding the distribution of competencies and skills across an organization---for example,
if employees who list hard skills typically do not list softer skills as their core competencies---may
help to provide insights into how businesses perform \citep{chin2020}, how effective projects are 
managed \citep{carvalho2015}, or how organizational or educational training 
programs are conceived \citep{robles2012,laker2011}.
In our analysis, we are thus interested in exploring if hard and softer skills are listed in homophilic 
patterns. To that end, we coded each competency as either ``hard skills'' or ``not hard skills'' and 
use the term {\tt b2nodematch("hardskill", [formulation] = [value])}, where {\tt [formulation]} may be 
replaced by {\tt alpha} for the node-centric formulation or {\tt beta} for the edge-centric 
formulation and {\tt [value]} may be replaced by a value from 0 to 1. 
The {\tt b2nodematch} parameter estimated here works analogously to the 
{\tt b1nodematch} equations presented in Section~\ref{sec:weightedNodematch},
but centers on the ``events'' mode---in this case, competencies. 
In addition, we include the {\tt edges} term, which controls the overall density of the network and is
useful in much the same way that it is typically useful to include an intercept term in any logistic 
regression model. 
In addition, we control for the effect of tenure of the employees, via {\tt b1cov("tenure")}, and the 
effect of gender, via {\tt b1factor("gender")}.

\begin{figure}[tb]
  \begin{center}
  \includegraphics[width = 0.3\textwidth]{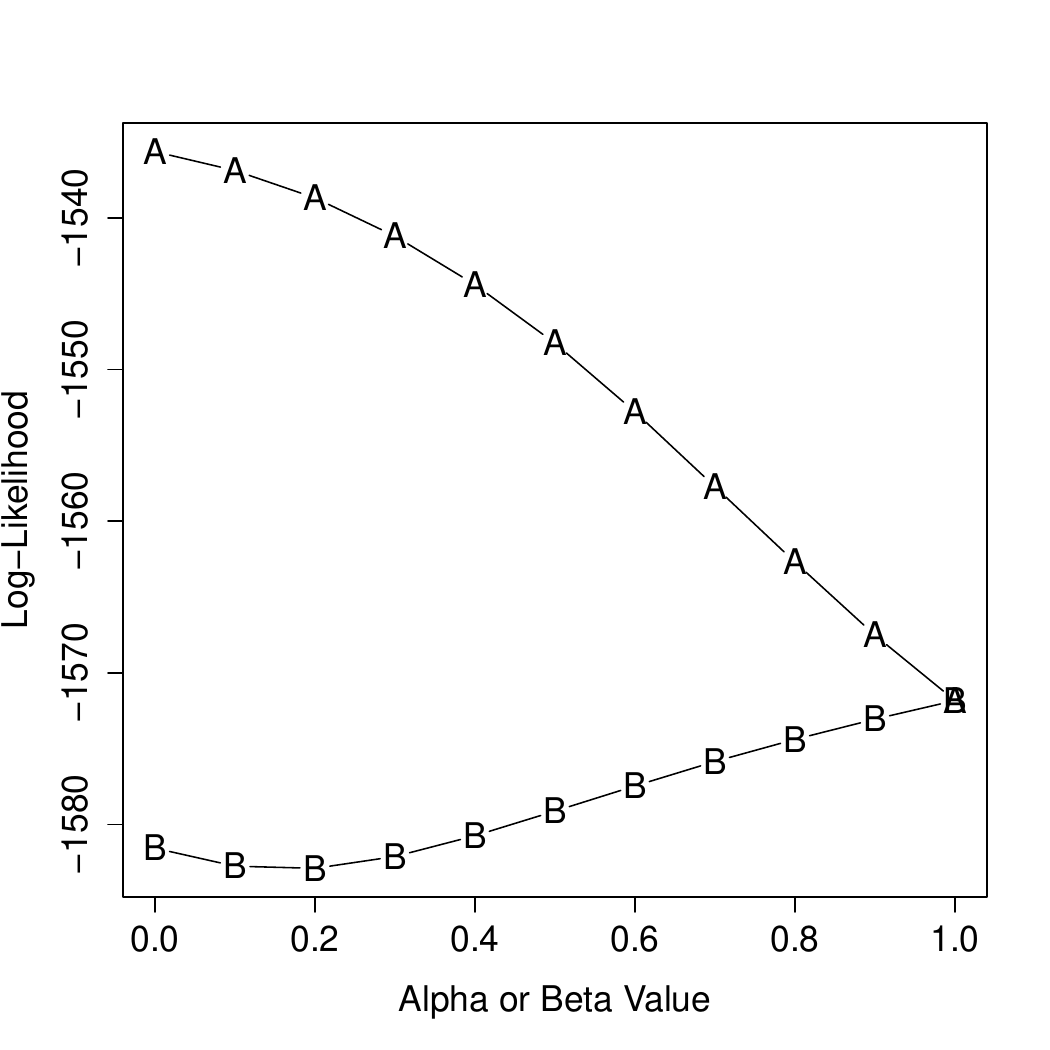}  
  \includegraphics[width = 0.3\textwidth]{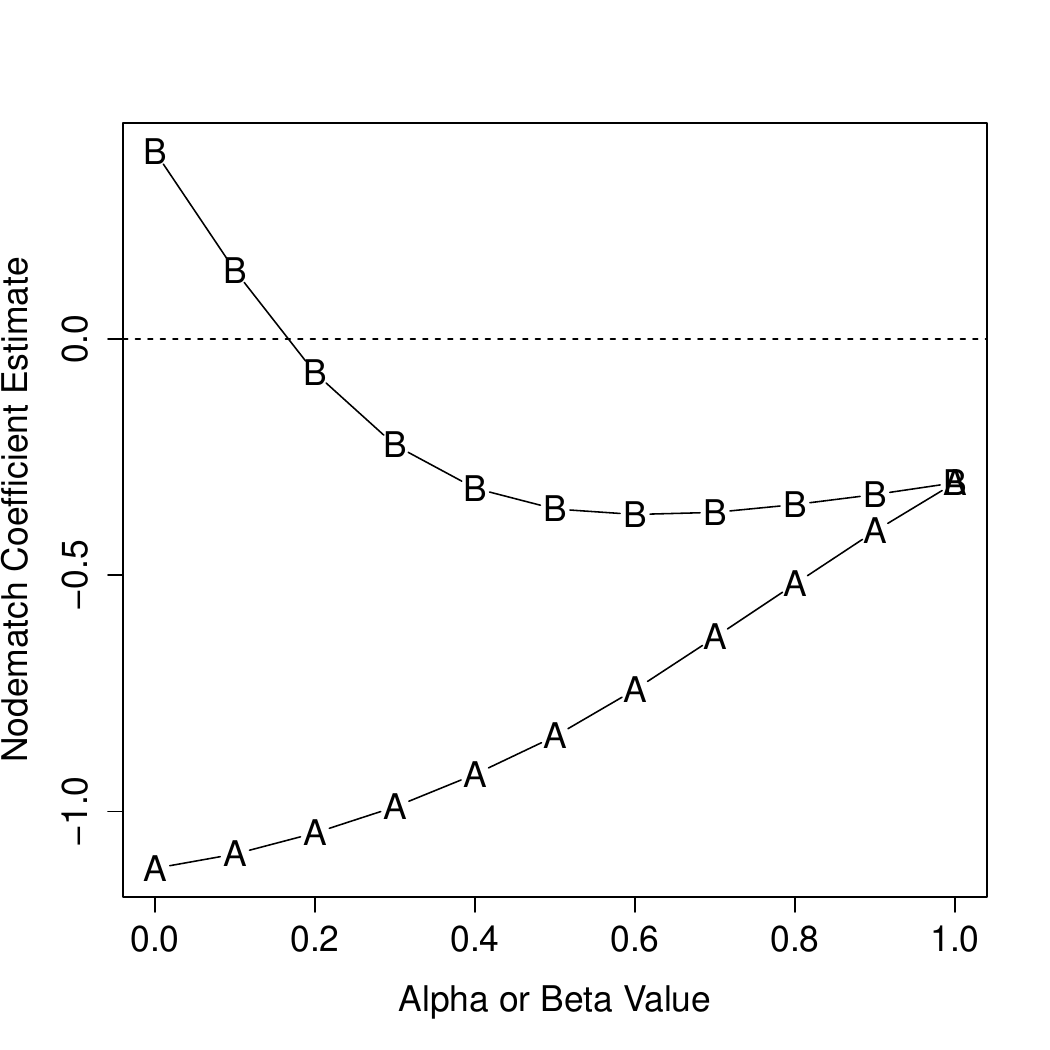}  
  \includegraphics[width = 0.3\textwidth]{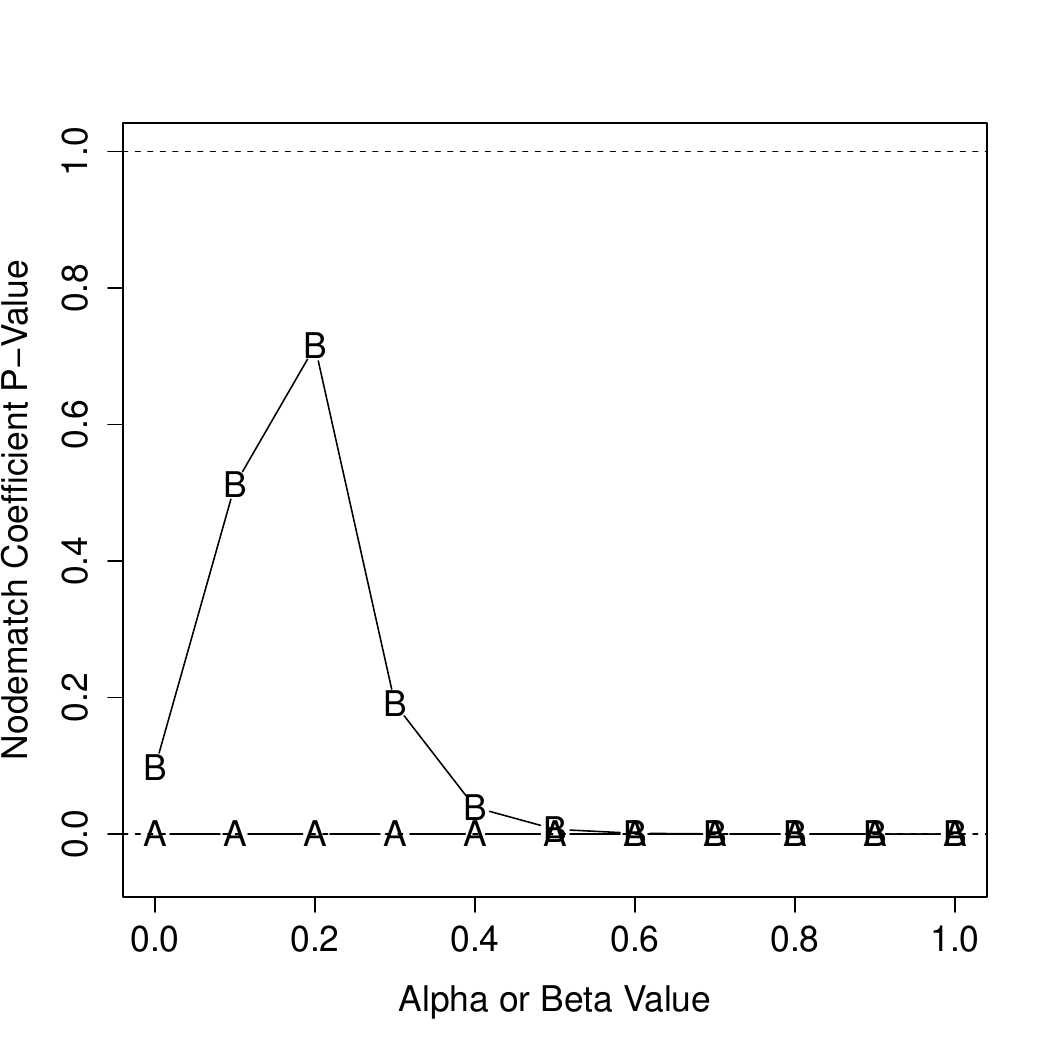}  
  \end{center}
  \caption{Left to right are (a) the approximate maximum log-likelihood values, (b) the coefficient estimates for the hardskill homophily term, and (c) the p-values associated with the hypothesis tests that the coefficients are zero, respectively, as a function of $\alpha$ (the A's) and $\beta$ (the B's).}
\label{fig:CompetencyLikelihoods}
\end{figure}

We approached the selection of $\alpha$ or $\beta$ values using a profile likelihood approach, thus fitting 
separate models for each parameter over a grid of points from 0 to 1. All models converged without 
issues. As is evident from the visualization in Figure~\ref{fig:CompetencyLikelihoods}(a),
the estimated profile likelihood 
suggests an optimal value of alpha=0. 
This means that homophily is measured by how many pairs of competences with the 
same categorization (hard vs.\ not) are found in at least one employee; absent a substantive
reason to prefer a particular homophily statistic, this version might be considered the
best choice based on its profile likelihood.
Table~\ref{tab:competency} presents the estimated parameters for the $\alpha = 0$ model.

\begin{table}[tb] \caption{
Parameter estimates and standard errors for the competency dataset, 
along with {\tt ergm} code used to obtain them.  Statistical significance at the 0.05, 0.001, and 0.0001 level is indicated by *, **, and ***, respectively.  }  
  \centering
  \begin{tabularx}{\textwidth}{l l rl} \hline\hline
  & {\bf Statistic }  &  {\bf {Parameter estimate} (S.E.)}\\  [0.5ex]
  \hline \hline

  & edges  &  \hspace {8 mm} $-1.936$ (0.152)***\textcolor{white}{}\\  [0.5ex]  
  & b1cov.tenure &  \hspace {8 mm}  $-0.001$ (0.047)\textcolor{white}{***}\\  [0.5ex]  
  & b1factor.gender.Male &  \hspace {8 mm}  $0.069$ (0.120)\textcolor{white}{***} \\  [0.5ex]  
  & b2nodematch.hardskill &  \hspace {8 mm} $-1.121$ (0.123)*** \\  [0.5ex]  
  \hline \hline 
\multicolumn{3}{l}{Code in {\tt ergm}:} \\
\multicolumn{3}{l}{
{\tt ergm(formula = net\_tm $\sim$ edges + b1cov("tenure") + }
}  \\ 
\multicolumn{3}{l}{
{\tt \quad b1factor("gender") + b2nodematch("hardskill", alpha = 0),}
} \\
\multicolumn{3}{l}{
{\tt \quad control = control.ergm(MCMC.samplesize = 15000, seed = 123456))}
} \\
\hline\hline
  \end{tabularx} 
  \label{tab:competency}
\end{table}

Our model leads to an estimated negative homophily effect: It is less likely that two competencies
sharing the same soft/hard designation will be listed by the same employee than would be
predicted by chance, after accounting for the observed overall density and the differential numbers of
competencies listed by people of different genders and tenures.
This effect persists when alternative model terms are considered.  For instance, we tried 
correcting for each individual competency's tendency to be named by employees, which is
accomplished by adding the {\tt b2sociality} model term.  In this case, the model with the highest
estimated log-likelihood was the $\alpha=0.5$ model, corresponding to a statistic in which
homophily is measured by summing, for each pair of matching competencies, the square root
of the number of employees who named both.  In this model and all of the models containing
{\tt b2sociality}, the homophily effect was statistically significantly negative.
We also duplicated the model terms shown in Table~\ref{tab:competency} but using the {\tt diff = TRUE}
option with the {\tt b2nodematch} term.  This modification implements differential homophily,
replacing the single statistic in Equation~(\ref{eqn:alpha}) by two statistics of the 
form~(\ref{eqn:alphaDifferential}),
one for $v=0$ (soft skills) and one for $v=1$ (hard skills).  The results revealed no interesting
qualitative difference in the two homophily effects nor any deviation from the overall results shown in
Figure~\ref{fig:CompetencyLikelihoods}.

In this case, we explored the full range of homophily parameters and
described the social implications post-hoc.  
Social 
scientists might prefer to let theory and hypothesis drive the construction of
statistical models.
We give an example for such an approach in our next illustration.

\subsection{Fortune 500 directorates}
\label{sec:directorates}

The role of gender in 
corporate governance and social elite cohesion is the subject of long-standing yet 
contemporary debates \citep{sidhu2021, sojo2016, kesner1988}.
Management scholars and sociologists alike have examined the formation and outcomes of gender 
diversity on corporate boards. For instance, \citet{seebeck2021} find that gender 
diversity on boards is positively related to the disclosure of corporate risks. \cite{kim2016}
find that diversified corporate boards improve firm value through additional expertise and unique 
skills. Other studies find that overall corporate performance tends to improve as a consequence of 
gender diversity on boards \citep{ohagan2017,erhardt2003},  solidifying the emerging consensus 
that females ascending ``through the ceilings of upper-middle management into strategy-making 
roles in the highest echelons, has a net benefit for firms in the long run'' \citep{jeong2017}.
Given these net benefits, it is perhaps surprising that, even today, data ``unequivocally show
that around the world, men hold the vast majority of corporate directorships and women are
starkly underrepresented'' \citep{kirsch2018}.

Here, we propose that, in addition to the well-known proverbial glass ceiling that prevents women
from rising further in the ranks of a company after achieving a certain threshold in their career,
there is also a glass door.  This effect is in evidence
after women have already obtained leadership positions, such 
as serving on a firm's board of directors, serving to ease further integration by women
onto these boards. 
Analyzing 
this horizontal integration contributes to research on the role of gender in board compositions and 
allows for novel insights into the phenomenon of social elite cohesion 
\citep{davis2003, park2013}. We will now demonstrate that the reformulated nodematch effects we 
propose in this paper allow for a nuanced analysis of the role of gender homophily, which helps to 
explain how women already present on boards may help to mitigate the effect of a proverbial glass 
door.

To that end, we take an edge-centric view of homophily.  That is, given a particular edge between
a female and a particular board, how likely is it that other females serve on the same board?
This is opposed to the node-centered view, which simply asks whether a given pair of females
is more or less likely to sit on a board together than pure chance might predict.
These two views coincide when $\alpha=\beta=1$, but
for values of $\alpha$ or $\beta$ strictly less than 1, the use of $\beta$ seems more appropriate
for this application.  Furthermore, the test for the glass door effect suggests a low value of $\beta$;
in the extreme when $\beta=0$, the \texttt{nodematch} term is the count, over all female-board 
combinations, of how many have at least one additional female on the same board.
The value $\beta=(\alpha=)1$, on the other hand, suggests that this glass door is opened much
more widely as the number of women increases, with the effect increasing quadratically in
the number of women.  We argue that such
a measure of homophily is inappropriate here; instead, our hypothesis is that
the first women serving on the board of a given firm somewhat normalize the
idea of female directors serving on the boards of said firms; 
after a few female directors, however, this effect decreases, as the notion of female 
directors serving the firm in an equal capacity has already been established.
In other words, based on wholly substantive considerations, we believe that homophily in
this case is captured most appropriately by a $\beta$ value that is small but strictly positive.

\begin{figure}[tb]
  \begin{center}
  \includegraphics[height = 2in, width = 2.5in]{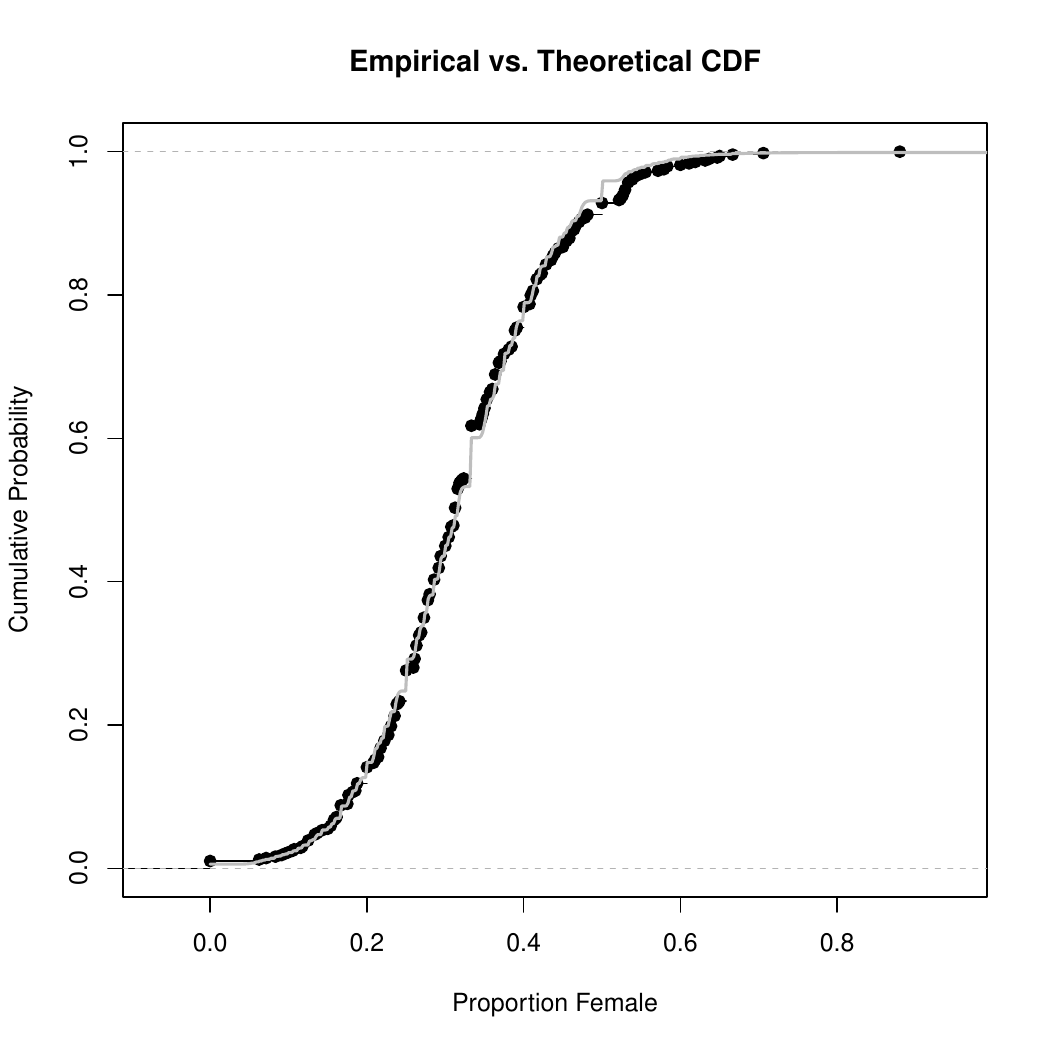}
  \end{center}
  \caption{Comparison of empirical (black dots) with theoretical (gray lines) cumulative distribution
  functions under the assumption of no gender-based 
  homophily for female proportions on 489 boards of directors.
   }
  \label{fig:CDFs}
\end{figure}

To estimate the gender homophily effect in practice, we analyze a network of
directorate memberships
among Fortune 500 companies in 2022.  Our sample consists of $n_1=489$ firms and
$n_2=8606$ directors. Here, directors are affiliated with different firms,
constituting a two-mode structure.
Preliminary exploration suggests that
if a nonzero gender homophily effect exists, it might be subtle. 
Figure \ref{fig:CDFs} depicts the empirical cumulative distribution
of the 489 values of $\hat p_F$, the sample proportion of females, among all of the boards.
If gender is irrelevant in board composition, we might expect that this distribution roughly
mimics what would occur if every board's gender makeup were binomially distributed with
parameters equal to the observed size of the board and the overall fraction,
$2741/8606$, of females among all directors.  As shown in Figure~\ref{fig:CDFs}, these two 
distribution functions are quite close.  

\begin{figure}[tb]
  \begin{center}
  \includegraphics[height = 2in, width = 2.5in]{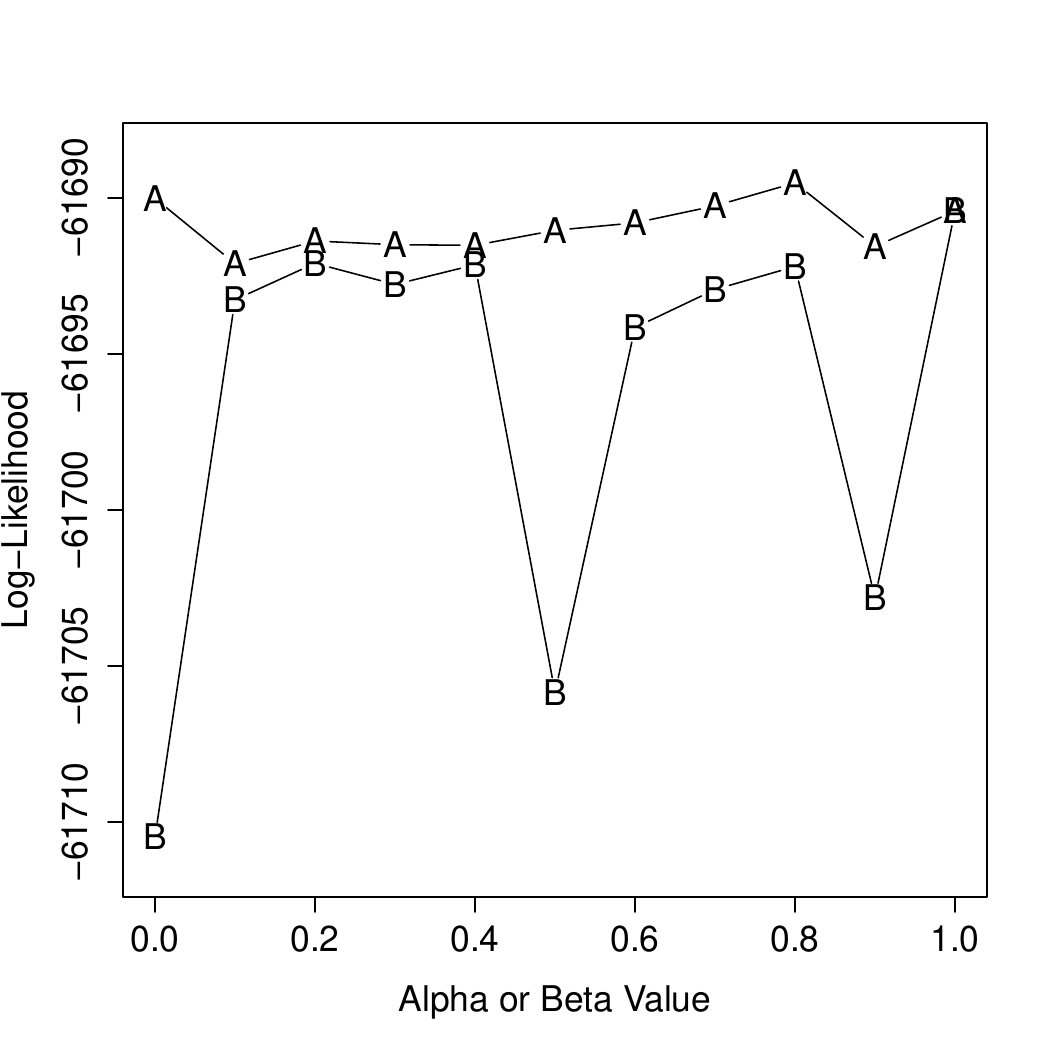}
  \includegraphics[height = 2in, width = 2.5in]{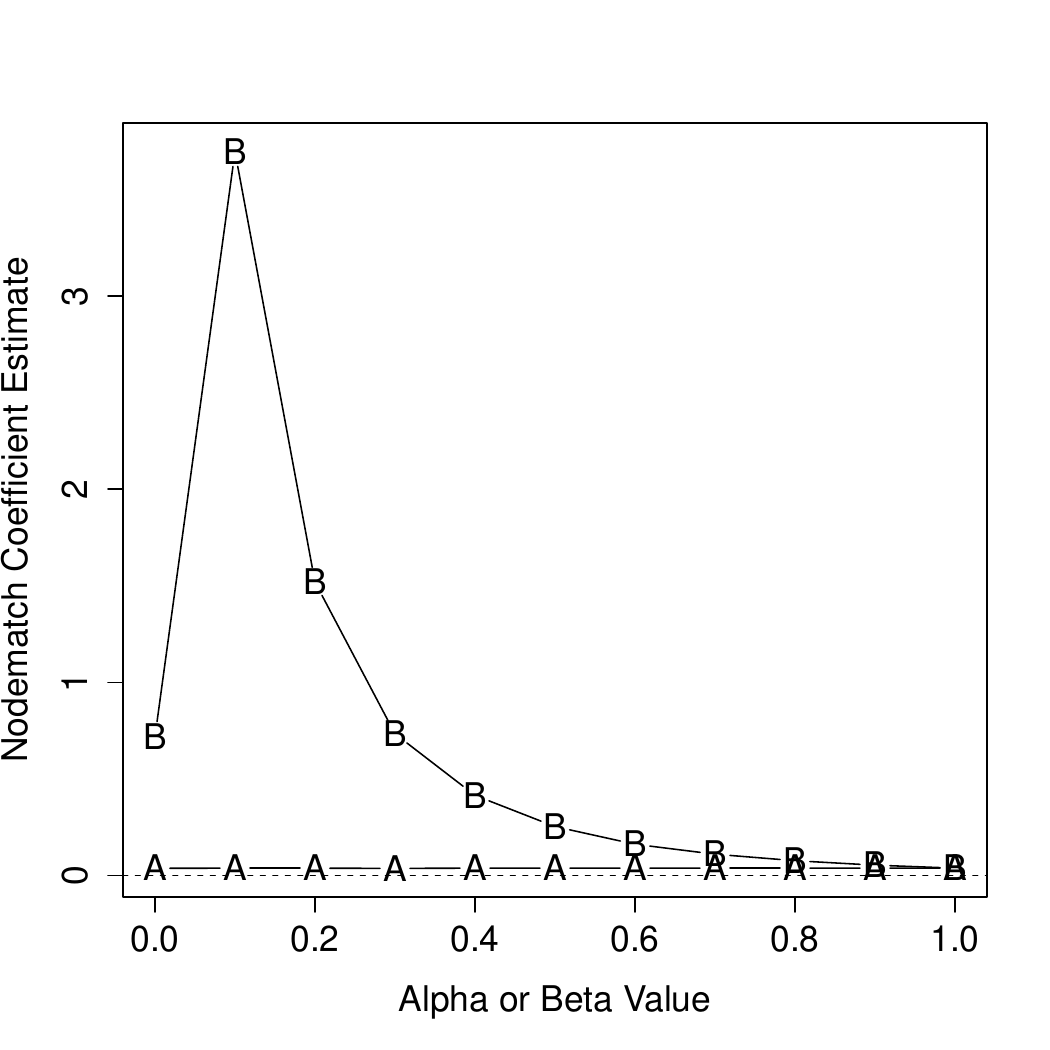}
  \end{center}
  \caption{Shown as a function of $\alpha$ (the A's) and $\beta$ (the B's) are (a) at left, the MCMC-
  based maximum log-likelihood approximations, each with an estimated standard deviation of 1;
  and (b) at right, the coefficient estimates for the female homophily term.}
  \label{fig:DirectorateEstimates}
\end{figure}

Additional exploratory analysis of the Fortune 500 dataset provides another reason
to prefer a $\beta$-based measure of homophily rather than an $\alpha$-based measure,
once we rule out the $\alpha=1$ idea for substantive reasons:  In this dataset, nearly all pairs
of females who serve on any board together 
serve on exactly one board together.  This fact is revealed if we compare the value of the
\texttt{nodematch(alpha = 0)} statistic, $10{,}582$, to the 
\texttt{nodematch(alpha = 1)} statistic, $10{,}592$.  In other words, the
$\alpha$-based homophily effects for this dataset are essentially all equivalent.
In contrast, the female
\texttt{nodematch(beta = 0)} statistic equals
only 1562, suggesting that this dataset has nuanced information regarding
the propensity for females serving on a board to predict whether other 
females also serve on that board and validating our decision to focus on a low $\beta$
value to address our substantive question about gender homophily on boards of directors.
Interestingly, a profile likelihood approach does not suggest a clear model choice in this case,
as shown in Figure~\ref{fig:DirectorateEstimates}(a).

Figure~\ref{fig:DirectorateEstimates}(a) shows the different parameter estimates produced by different $\alpha$ and $\beta$ values from 0 to 1.  
All estimates are significant at $p<0.001$ except the $\beta=0$ case, which was also
considerably more difficult than others for
{\tt ergm}'s MCMC-based estimation procedure to fit.  Indeed, the fitting procedure
hinted at near-collinearity in the $\beta=0$ case, providing another reason to
select a strictly positive $\beta$ value.

Estimated coefficients for a model using $\beta=0.1$ that also corrects for nodal
covariates including
gender (for directors) and assets and industry sector (for companies) are shown in 
Table~\ref{tab:coexpansiondiff}.  This model also corrects for the overall propensity
for directors to join multiple boards, via {\tt b2star(2)} and {\tt b2degree(1)}, without
regard to gender.  The ability to correct for such effects, in a regression-like
setting, for nodes from each of the two modes is one of the virtues of the methods
presented here that retain the full bipartite structure of the dataset.

Differentiating between the effects of gender homophily for men and women 
shows that the estimated effect is stronger for women than for their male counterparts.
However, this difference is not statistically significant; in the fitted model summarized by
Table~\ref{tab:coexpansiondiff}, the difference between the female
and male homophily coefficient 
estimates equals $0.85$ whereas the estimated standard error of this difference is
$0.94$.  Such contrasts may be analyzed using standard techniques because the
fitted model produced by the {\tt ergm} package includes an estimated covariance matrix.

\begin{table}[tb] \caption{
Parameter estimates and standard errors for the directorates dataset,
using $\beta=0.1$ for the homophily terms, along with {\tt ergm} code 
used to obtain them. 
Statistical significance at the 0.0001 level is indicated by ***.  }  
  \centering
  \begin{tabularx}{\textwidth}{l l rl} \hline\hline
  & {\bf Statistic }  &  {\bf {Parameter estimate} (S.E.)}\\  [0.5ex]
  \hline \hline  
  & edges & \hspace {8 mm} $-7.61$ (0.59)*** \\  [0.5ex]  
  & b2star2 & \hspace {8 mm} $-2.47$ (0.17)*** \\  [0.5ex]  
  & b2degree(1) &  \hspace {8 mm} $3.48$ (0.17)*** \\  [0.5ex]  
  &b1cov.log10\_total\_assets &  \hspace {8 mm} $0.097$ (0.018)*** \\  [0.5ex]  
  & b1factor.industry\_sector.2 &  \hspace {8 mm} $-0.030$ (0.035)\textcolor{white}{***} \\  [0.5ex]  
  & b1factor.industry\_sector.3 &  \hspace {8 mm} $0.039$ (0.053)\textcolor{white}{***} \\  [0.5ex]  
  & b1factor.industry\_sector.4 &  \hspace {8 mm} $0.097$ (0.051)\textcolor{white}{***} \\  [0.5ex]  
  & b1factor.industry\_sector.5 &  \hspace {8 mm} $-0.086$ (0.034)*\textcolor{white}{**} \\  [0.5ex]  
  & b1factor.industry\_sector.6 &  \hspace {8 mm} $0.030$ (0.046)\textcolor{white}{***} \\  [0.5ex]  
  & b2nodematch.gender.Female &  \hspace {8 mm} $3.75$ (0.67)*** \\  [0.5ex]  
  & b2nodematch.gender.Male &  \hspace {8 mm} $2.90$ (0.72)*** \\  [0.5ex]  
  & b2factor.gender.Male &  \hspace {8 mm} $0.12$ (0.66)\textcolor{white}{***} \\  [0.5ex]  
  & b2cov.age &  \hspace {8 mm} $0.045$ (0.004)*** \\  [0.5ex]  
  \hline \hline 

\multicolumn{3}{l}{Code in {\tt ergm}:} \\
\multicolumn{3}{l}{
{\tt ergm(Directorships $\sim$ edges + b2star(2) +  b2degree(1) }
} \\ 
\multicolumn{3}{l}{
{\tt \quad + b1cov("log10\_total\_assets") + b1factor("industry\_sector")}
} \\
\multicolumn{3}{l}{
{\tt \quad +  b2nodematch("gender", beta = 0.1, diff = TRUE)}
} \\
\multicolumn{3}{l}{
{\tt \quad +  b2factor("gender") + b2cov("age")}
} \\
\hline\hline
  \end{tabularx} 
  \label{tab:coexpansiondiff}
\end{table}

Our findings contribute to ongoing discussions about women in the boardroom
\citep{kirsch2018, mah2022}, illuminating how a different perspective on homophily that takes 
bipartite structures into account leads to novel approaches to long-standing questions.

\section{Discussion}
\label{sec:discussion}

Apart from the statistical benefits, such as increased modeling flexibility and improved model fit,
the reformulated nodematch effects presented here offer 
exciting avenues for future research for scholars interested in social network analysis.
Most notably, the reformulated effects enable questions that do not simply employ a one-mode
perspective on homophily after projecting from a two-mode network. Rather, they are
built with the properties of two-mode networks in mind, allowing for the differentiation of
two fundamentally different theoretical constructs: Homophily based on enumerating
paths that connect a particular node pair with the same attributes, versus homophily based on counting
completed two-paths with matching end nodes that include a particular tie.  Only in the case where
these counts contribute linearly to the corresponding homophily statistics are these two views equivalent.

We present two real-world applications to
demonstrate both of these perspectives and find contrasting results: For the first network,
in which employees are connected through shared competences they were asked to list,
the node-centric view dramatically alters the
estimated models' statistical properties and therefore the substantive conclusions we derive from
analyzing the observed network.
For the second network, which depicts
board memberships among Fortune 500 companies, the edge-centric reformulation 
provides insight into pressing questions that so far had remained
unanswered. These two contrasting results underscore that the reformulated effects we
present in this article allow for theory-building and subsequent empirical testing from a
novel yet natural two-mode perspective.

The methods illustrated here are available in the {\tt ergm} package for {\tt R}, where they may be
used in conjunction with a wide range of other statistics.  The specific statistics
included in an ERGM will depend on the modeling aims particular to the situation and,
as always, the interpretation of coefficients from a fitted model must be done in the
presence of all other model terms.  General treatments of ERGMs \citep[e.g.,][]{robins2007ait} or articles specific to the bipartite network setting
\citep[e.g.,][]{wang2009,wang2013} provide insight into additional model terms that may be of interest.

\section{Acknowledgments}
This work is supported by NIH grant R01-GM083603-01.
The authors are grateful to Shweta Bansal of Georgetown University,
who was involved in early conversations
about this manuscript and who supplied motivating examples not actually used here.
The data on competencies were graciously provided by Andrew Parker
of Durham University.

\appendix

\section{Change Statistics}
\label{sec:changestats}

In addition to providing insight on model coefficients as explained in Section~\ref{sec:interpretation}, examination of change statistics may sometimes identify 
possible degeneracy problems with an ERGM:  If a change statistic can grow exceedingly
large in the positive or negative direction, then a corresponding $\theta$ coefficient of
the same or opposite sign can virtually assure the presence or absence of an edge,
and this type of behavior can become self-reinforcing.  We do not delve into details of the
degeneracy issue here, referring interested readers instead to 
\citet{handcock2003} and \citet{schweinberger2011}.  

This section presents change statistics for both the node-centered ($\alpha$-based)
and edge-centered ($\beta$-based) homophily statistics introduced in this article. 
First, fix a node pair $(i, k)$, where $i$ belongs to mode~1, $k$ belongs to mode~2, and $c$ is a categorical nodal attribute measured on the mode~1 nodes.
We will focus on the {\tt b1nodematch} change statistics, though the arguments below
may easily be adapted to the {\tt b2nodematch} case.

The corresponding change statistics are obtained by calculating the difference 
between the contributions to the statistic with $y_{ik} = 1$ and $y_{ik} = 0$. 
For the node-centric ($\alpha$-based) version of the homophily statistic, 
Equation~(\ref{eqn:alpha}) leads to
\begin{equation*}
  \delta^{[\alpha]}(\vec y)_{ik} = 
  \mathop{\sum}_{\stackrel{ j\ne i :}{\mbox{\scriptsize $c_j=c_i$}}}
  \left[ \sum_{k' \neq k} 
  y_{ik'} y_{jk'} + y_{jk} \right]^\alpha 
  -  
  \mathop{\sum}_{\stackrel{ j\ne i :}{\mbox{\scriptsize $c_j=c_i$}}}
  \left[ \sum_{k' \neq k} 
  y_{ik'} y_{jk'} \right]^\alpha,
\end{equation*}
where the factor of 1/2 from Equation~(\ref{eqn:alpha}) is unnecessary because
here we are not double-counting the contribution of $y_{ik}$ to the overall
statistic.
This expression may be rewritten as
 \begin{equation}\label{eqn:changestat_alpha} 
  \delta^{[\alpha]}(\vec y)_{ik} = 
  \mathop{\sum}_{\stackrel{ j\ne i :}{\mbox{\scriptsize $c_j=c_i$}}}
  y_{jk}
  \left\{[t_{\vec y}(i, j, k) + 1]^\alpha - t_{\vec y}(i, j, k)^\alpha \right\},
\end{equation}
where $t_{\vec y}(i, j, k)$ is the number of two-paths in $\vec y$ from $i$ to $j$ not passing via $k$.

For the edge-centered ($\beta$-based) version of the homophily statistic, let
us choose a particular $1\le i\le n_1$ and $n_1+1\le k \le n$.  
Define $u_{\vec y}(i,k)$ to be the number of distinct values $j\ne i$ such that
$y_{jk}=1$ and $c_j=c_i$, i.e., the number of edges to $k$ from nodes matching $i$.  
Then if $y_{ik}=1$, Equation~(\ref{eqn:beta}) may be split into three terms,
\begin{equation}\label{eqn:ThreeTerms}
  \mbox{b1nodematch}(\vec y; \beta) = \frac12 u_{\vec y}(i,k)
  + \frac12 
  \mathop{\sum}_{\stackrel{ j\ne i :}{\mbox{\scriptsize $c_j=c_i$}}}
  [u_{\vec y}(i,k)] ^ \beta
  + C,
\end{equation}
where $C$ does not depend on $y_{ik}$.  When we instead take $y_{ik}=0$, the
first term above disappears, each summand in the second term is
$[u_{\vec y}(i,k)-1]^\beta$, and of course $C$ does not change.  
To find the change statistic, therefore, we subtract---after noticing that
the sum in (\ref{eqn:ThreeTerms}) has exactly $u_{\vec y}(i,k)$ terms---to obtain
\begin{eqnarray}  \label{eqn:changestat_beta}
  \delta^{[\beta]}(\vec y)_{ik} & = \frac12 [u_{\vec y}(i,k)]^\beta + \frac12 u_{\vec y}(i,k)[u_{\vec y}(i,k)]^\beta - \frac12 u_{\vec y}(i,k)[u_{\vec y}(i,k)-1]^\beta \\ \nonumber
& = \frac12 \left[(1+u)u^\beta - u(u-1)^\beta \right],
\end{eqnarray}
where the last line uses $u\equiv u_{\vec y}(i,k)$ for simplicity of notation.

Considering the extreme case in which $\alpha$ or $\beta$ equals 0, we see 
from Equation~(\ref{eqn:changestat_beta}) that when $\beta=0$,
the change statistic
cannot be larger than 1/2, which means that degeneracy behavior is unlikely.  When
$\alpha=0$, on the other hand, it takes a bit of thought to see that the summand in Equation~(\ref{eqn:changestat_alpha}) is zero unless
both $t(i, j, k; \vec y)=0$ and $y_{jk}=1$.  Thus, the value of 
$\delta^{[\alpha]}(\vec y)_{ik}$ could theoretically be as large as one half the
number of mode-1 nodes minus one, yet such a large value would not
occur repeatedly as the number of edges increases toward a complete network
or decreases toward an empty network.  Thus, we do not expect degeneracy behavior when either $\alpha=0$ or $\beta=0$.

\section{Curved Exponential Families}
\label{sec:curvedEF}

\citet{hunter2006iic} 
demonstrated the utility of applying the statistical concept of
curved exponential-family models 
\citep{efron1975dcs,efron1978tgo} 
to the modeling of networks.  In a curved ERGM,
the standard ERGM of Equation~(\ref{basicERGM}) is modified by 
assuming that the linear combination of the $s_i(\vec y)$ statistics is not defined 
by the $\vec\theta$ parameters directly, but rather by some function $\vec\eta(\vec\theta)$.
Thus, Equation~(\ref{basicERGM}) is replaced by
  \begin{equation}
  \label{cergm}
  P(\vec Y =  \vec y) = \frac{\exp\{ \sum_{i=1}^p 
  \eta_i(\vec\theta) s_i(\vec y) \}}{\kappa[\vec\eta(\vec\theta)]},
  \end{equation}
where $\vec s(\vec y)$ is a $p$-dimensional vector of network statistics on $\vec y$,
as before, but now $\vec\theta$, the parameter vector of interest 
is $q$-dimensional for some $q<p$.  The two vectors are related by the $p$-dimensional
vector $\vec\eta$, which is assumed to be a function of $\vec\theta$.
As usual, statistical estimation focuses on the maximum likelihood estimator of $\vec\theta$,
\[
 \hat{\vec\theta} = \mbox{arg max}_{\vec\theta \in \mathbb{R}^q}
\frac{\exp\{ \sum_{i=1}^p \eta_i(\vec\theta) s_i(\vec y) \}}{\kappa[\vec\eta(\vec\theta)]}.
 \]
Here, we demonstrate that the statistics defined by
Equations~(\ref{eqn:alpha}) and~(\ref{eqn:beta})
may be rewritten in curved exponential-family form, where $\alpha$ or $\beta$
plays the role of the $\theta$ parameter of interest seen in Equation~(\ref{cergm}).

Recall that $n_2$ is the number of mode~2 nodes in the network.  Let us define, for
$0\le i\le n_2$, the mode-1 matching dyadwise shared partner statistic
$\mbox{b1MDSP}_i(\vec y)$ to equal the number of matching pairs of mode-1 nodes
that have exactly $i$ common (mode-2) neighbors.  Then we may
rewrite Equation~(\ref{eqn:alpha}) as
\begin{equation}
  \mbox{b1nodematch}(y, \alpha) = 
  \sum_{i=1}^{n_2}
  {(i^\alpha)\times \mbox{b1MDSP}_i(\vec y)}.
  \label{eqn:alpha_cergm}
\end{equation}
Similarly, for any $0\le i\le n_1-1$, if $\mbox{b1MESP}_i(\vec y)$ is the number of edges in $\vec y$ 
that form a ``mode-1 matching two-path'' with exactly $i$ other edges---i.e., that are contained 
in exactly 
$i$ two-paths connecting two mode-1 nodes that match on the attribute of interest---then
we may rewrite Equation~(\ref{eqn:beta}) as
\begin{equation}
  \mbox{b1nodematch}(y, \beta) = \frac12
  \sum_{i=1}^{n_1-1}
  {(i^\beta)\times\mbox{b1MESP}_i(y)}.
  \label{eqn:beta_cergm}
\end{equation}

The $\mbox{b1MDSP}$ and $\mbox{b1MESP}$ statistics may be viewed as roughly 
analogous to the DP and EP statistics
that give rise to the geometrically weighted dyadwise shared partner (GWDSP) and 
edgewise shared partner (GWESP) statistics
explained in \citet{hunter2007cef}.  
In fact, the b1MDSP statistics are precisely a homophily-based version of the DP 
(dyadic shared partner) statistics in \citet[][Equation (26)]{hunter2007}, which demonstrates
that when $\beta=0$, our Equation~(\ref{eqn:beta_cergm}) is a homophily-based special
case of the alternating $k$-twopath statistic of \citet{robins2007rdi}.  However, the analogy between
b1MESP and EP statistics is not quite as direct.  Indeed, the EP statistics of 
\citet{hunter2007}---and thus the alternating $k$-triangle statistic of \citet{robins2007rdi}---are
meaningless in a bipartite context since triangles are impossible.

If we view $\alpha$ as a fixed constant in Equation~(\ref{eqn:alpha_cergm}), 
then the coefficients $i^\alpha$---which are the parameters
$\eta_i(\theta)$ seen in Equation~(\ref{cergm})---are fixed.  Therefore the ``curvature''
is lost, and the model is a standard (non-curved) ERGM.  On the other hand,
if we view $\alpha$ as an unknown parameter to be estimated, then the complications
of curved exponential families, as explained in 
\citet{hunter2006iic}, 
arise.
The same arguments are true of the $\beta$ in Equation~(\ref{eqn:beta_cergm}).

\bibliographystyle{imsart-nameyear}
\bibliography{bibliography}

\end{document}